\documentclass{aa}

\usepackage{graphicx}
\usepackage{txfonts}
\usepackage{textcomp}
\usepackage{natbib}

\bibpunct{(}{)}{;}{a}{}{,} 
\begin{document}

   \title{Measurement of the solar system acceleration using the Earth scale factor}


   \author{O. Titov\inst{1}
          \and
          H. Kr\'asn\'a\inst{2}
          }

   \institute{Geoscience Australia, PO Box 378, Canberra, 2601, Australia\\
              \email{Oleg.Titov@ga.gov.au}
         \and
             Technische Universit\"at Wien, Department of Geodesy and Geoinformation,
             Gusshausstrasse 27-29, A-1040 Vienna, Austria\\
             }

   \date{Received 8 September 2017 / Accepted 30 November 2017}


  \abstract
   {}
   {We propose an alternative method to detect the secular aberration drift induced by the solar system acceleration due to the attraction to the Galaxy centre. This method is free of the individual radio source proper motion caused by intrinsic structure variation.}
   {We developed a procedure to estimate the scale factor directly from very long baseline interferometry (VLBI) data analysis in a source-wise mode within a global solution. The scale factor is estimated for each reference radio source individually as a function of astrometric coordinates (right ascension and declination). This approach splits the systematic dipole effect and uncorrelated motions on the level of observational parameters.}
   {We processed VLBI observations from 1979.7 to 2016.5 to obtain the scale factor estimates for more than 4,000 reference radio sources. We show that the estimates highlight a dipole systematics aligned with the direction to the centre of the Galaxy. With this method we obtained a Galactocentric acceleration vector with an amplitude of 5.2 $\pm$ 0.2~\textmu as/yr and direction $\alpha_G = 281\degr \pm 3\degr$ and $\delta_G = -35\degr \pm 3\degr$.}
   {}

   \titlerunning{Solar system acceleration and the scale factor}
   \maketitle

%

\section{Introduction}

The solar system is rotating around the Galactic centre with a period of approximately 200~million years. The solar system barycentre shows, therefore, rotational acceleration in the direction of the Galactic centre of 5-6 microarcseconds of arc per year (\textmu as/yr) and this effect, the secular aberration drift (SAD), results in a systematic proper motion of reference radio sources around the sky.

The theoretical prediction of the dipole systematics (\citet{Fanselow83, Bastian95, Eubanks95, Gwinn97, Sovers98, Mignard02, Kovalevsky03, Kopeikin06}) was recently confirmed after the analysis of a global set of very long baseline interferometry (VLBI) data since 1979 (\citet{Titov11, Xu12, Xu13, Titov13, Titov16, MacMillan14}). The magnitude of the dipole systematics is consistent, whereas the estimate of the dipole direction varies mostly owing to the strong influence of the variability of the relativistic jets (Table~\ref{app_prop_motions}). The individual apparent proper motion of the reference radio sources may reach 0.5~millisecond of arc/yr (i.e. 100 times of the dipole effect).  Therefore, the exclusion of astrometrically unstable radio sources from the analysis is a highly labour-intensive process that yields marginal results with a formal error of 1~\textmu as/yr (e.g. \citet{Titov13}). In the meantime, \citet{Mignard12} showed that the Gaia mission will obtain the parameters of the Galactocentric acceleration with a formal error of 0.2 \textmu as/yr, i.e. five times better. Therefore, we are seeking methods to improve the VLBI data results.

Theoretically, the Galactocentric acceleration vector points to the centre of the Galaxy, which is supposed to be the location of the compact radio source Sagittarius A* with the estimated coordinates $\alpha_G$ = 267\degr for right ascension and $\delta_G$ = $-29$\degr for declination having the absolute position error of 0\farcs12 \citep{Reid04}. A comprehensive instruction to calculate the acceleration amplitude using the International Astronomical Union-recommended Galactic centre distance ($R_{gal} = 8.5$~kpc) with the Galaxy rotation velocity at $R_{gal}~( V_{gal} = 220$~km/s) and assuming the circular type of motion is given, for example in \citet{Kovalevsky03} or \citet{Titov11}. This results in a value
$A = |\boldsymbol{a}| = \frac{V_{gal}^2}{R_{gal}} = 1.85 \cdot 10^{-13}~\textrm{km/s}^2$, which can be converted to the acceleration vector $\boldsymbol{a}$ with the magnitude of 3.8~\textmu as/yr. There are dozens of papers devoted to revision of the parameters $R_{gal}$ and $V_{gal}$ based on different modern observation data and analysis strategies. For instance, \citet{Reid09} provided new meanings for $R_{gal} = 8.4 \pm 0.6$~kpc and $V_{gal} = 254 \pm 16$~km/s. The corresponding acceleration amplitude is $2.49 \cdot 10^{-13}~\textrm{km/s}^2$ and the amplitude $A$ of the dipole proper motion equals to 5.1~\textmu as/yr.
In one of the most recent publications, \citet{deGrijs16} recommended a downward adjustment of the $R_{gal}$ to 8.3~kpc and an upward revision of $V_{gal}$ in the range from 232~km/s to 266~km/s. The corresponding amplitude of the Galactocentric acceleration vector lies therefore within the range $(2.10 \div 2.76) \cdot10^{-13}~\textrm{km/s}^2$ followed by the expected amplitude of the dipole proper motion between 4.3 and 5.7~\textmu as/yr.

\citet{Titov11a} noted that the conventional equation of the geodetic VLBI group delay \citep[chap.~11]{iers10} could be altered to accommodate the Galactocentric acceleration. This modification provides a way to estimate the SAD either as a dipole systematics in proper motions or as a systematic effect in the VLBI scale. Section~\ref{Sec_basic_eq} recalls the corresponding mathematical equations and demonstrates the advantage of the latter option. In Section~\ref{Sec_analysis}, we present the results obtained by applying the new method to analysis of VLBI data during 1979.7-2016.5 together with the SAD estimates from global VLBI adjustments.

\begin{table*}
  \caption[]{Estimates of the SAD parameters from available publications if only a dipole component is estimated. The components of the acceleration vector are given with $a_1, a_2, a_3$, the amplitude with $A$, and $\alpha_G$, $\delta_G$ are the equatorial coordinates of the vector direction.}
     \label{app_prop_motions}

\begin{tabular}{lrrrrr}
        \hline
        \noalign{\smallskip}
   &\citet{Titov11}&    \citet{Xu12}&   \citet{Titov13} &\citet{MacMillan14}&   \citet{Titov16}\\
        \noalign{\smallskip}
        \hline
        \noalign{\smallskip}
$a_1$ [\textmu as/yr]&  $-0.7 \pm 0.8 $&$       -2.6 \pm 0.3$&$    -0.4 \pm 0.7 $&$ -0.3 \pm 0.3 $&$        0.2 \pm 0.8$ \\
$a_2$ [\textmu as/yr]&  $-5.9 \pm 0.9 $&$       -5.1 \pm 0.3 $&$        -5.7 \pm 0.8 $&$     -5.4 \pm 0.3 $&$        -3.3 \pm 0.8$\\
$a_3$ [\textmu as/yr]&  $-2.2 \pm 1.0 $&$       -1.1 \pm 0.4 $&$ -2.8 \pm 0.9 $&$ -1.1 \pm 0.5 $&$        -4.9 \pm 1.0$\\
$A$ [\textmu as/yr] &   $6.4 \pm 1.5 $&$        5.8 \pm 0.4 $&$ 6.4 \pm 1.1 $&$     5.6 \pm 0.4 $&$ 5.9 \pm 1.0$\\
$\alpha_G$ [\degr]   &  $263 \pm 11 $&$         243 \pm 4 $&$   266 \pm 7 $&$     267 \pm 3 $&$   273 \pm 13$\\
$\delta_G$ [\degr]   &  $-20\pm 12 $&$          -11 \pm 4 $&$   -26 \pm 7 $&$     -11 \pm 3 $&$   -56 \pm 9$\\
        \noalign{\smallskip}
        \hline
\end{tabular}

\end{table*}

\section{Basic equations}
\label{Sec_basic_eq}
The fundamental geodetic VLBI observable, the group delay $\tau_{group}$, is the first derivative of the phase with respect to angular frequency of the incoming signal. A group of eight channels within the frequency band from 8.1 to 8.9~GHz (known as X-band) measures the phases simultaneously to obtain the group delay by the least-squares adjustment. The conventional group delay model approximating the observed VLBI data is given by~\citet[chap.~11]{iers10} as

\begin{equation}\label{groupdelay_gcrs}
\tau_{group}=\frac{-\frac{\boldsymbol{b}\cdot\boldsymbol{s}}{\textrm{c}}\Big(1-\frac{2GM_{\odot}}{\textrm{c}^{2} R_{\oplus\odot}} -\frac{|\boldsymbol{V}_{\oplus}|^{2}}{2\textrm{c}^{2}}-\frac{\boldsymbol{V}_{\oplus}\cdot\boldsymbol{w}_{2}}{\textrm{c}^2}\Big) -\frac{\boldsymbol{V}_{\oplus}\cdot\boldsymbol{b}}{\textrm{c}^{2}}\Big(1+\frac{\boldsymbol{s}\cdot\boldsymbol{V}_{\oplus}}{2\textrm{c}}\Big)}
    {1+\frac{\boldsymbol{s}\cdot(\boldsymbol{V}_{\oplus}+\boldsymbol{w}_{2})}{\textrm{c}}}
,\end{equation}

where $\boldsymbol{b}$ is the vector of baseline $\boldsymbol{b} = \boldsymbol{r_2}-\boldsymbol{r_1}$, $\boldsymbol{s}$ is the barycentric (solar system barycentre) unit vector of radio source, $\boldsymbol{V}_{\oplus}$ is the barycentric velocity of the geocentre, $\boldsymbol{w}_{2}$ is the geocentric velocity of the second station, \textrm{c} is the speed of light, $G$ is the gravitational constant, $M_{\odot}$ is the mass of the Sun, and $R_{\oplus\odot}$ is the geocentric distance to the Sun.

The term $\frac{2GM_{\odot}}{\textrm{c}^{2} R_{\oplus\odot}}$ is related to the general relativity effect. The impact of $\boldsymbol{w}_{2}$ is small and may be ignored for sake of simplicity. After these alterations equation~(\ref{groupdelay_gcrs}) is given by

\begin{equation}\label{groupdelay_gcrs_simple}
\tau_{group}=\frac{-\frac{\boldsymbol{b}\cdot\boldsymbol{s}}{\textrm{c}}\Big(1 -\frac{|\boldsymbol{V}_{\oplus}|^{2}}{2\textrm{c}^{2}}\Big) -\frac{\boldsymbol{V}_{\oplus}\cdot\boldsymbol{b}}{\textrm{c}^{2}}\Big(1+\frac{\boldsymbol{s}\cdot\boldsymbol{V}_{\oplus}}{2\textrm{c}}\Big)}
    {1+\frac{\boldsymbol{s}\cdot\boldsymbol{V}_{\oplus}}{\textrm{c}}} .
\end{equation}

Equation~(\ref{groupdelay_gcrs_simple}) now comprises only the effects of special relativity of orders $\frac{\boldsymbol{V}_{\oplus}}{\textrm{c}}$ and $\frac{\boldsymbol{V}_{\oplus}^2}{\textrm{c}^2}$. Leaving only the terms of $\frac{\boldsymbol{V}_{\oplus}}{\textrm{c}}$ and using the Taylor series expansion for $(1+\frac{\boldsymbol{s}\cdot\boldsymbol{V}_{\oplus}}{\textrm{c}})^{-1}$, the equation~(\ref{groupdelay_gcrs_simple}) comes down to the main aberration effect

\begin{equation}\label{groupdelay_gcrs_main_aberr}
\begin{aligned}
\tau_{group}=&
    -\frac{\boldsymbol{b}\cdot\boldsymbol{s}}{\textrm{c}} \Bigg(1 -\frac{\boldsymbol{s}\cdot\boldsymbol{V}_{\oplus}}{\textrm{c}}\Bigg) -\frac{\boldsymbol{V}_{\oplus}\cdot\boldsymbol{b}}{\textrm{c}^{2}} =\\
    =&
    -\frac{\boldsymbol{b}\cdot\boldsymbol{s}}{\textrm{c}} + \frac{(\boldsymbol{b}\cdot\boldsymbol{s})}{\textrm{c}}\frac{(\boldsymbol{s}\cdot\boldsymbol{V}_{\oplus})}{\textrm{c}}
    -\frac{\boldsymbol{V}_{\oplus}\cdot\boldsymbol{b}}{\textrm{c}^{2}} =\\
    =& -\frac{\boldsymbol{b}\cdot\boldsymbol{s}}{\textrm{c}} + \frac{\boldsymbol{b}\cdot(\boldsymbol{s}(\boldsymbol{s}\cdot\boldsymbol{V}_{\oplus})-\boldsymbol{V}_{\oplus})}{\textrm{c}^{2}} =\\
    =&
    -\frac{\boldsymbol{b}\cdot\boldsymbol{s}}{\textrm{c}} +
    \frac{\boldsymbol{b}\cdot(\boldsymbol{s}\times(\boldsymbol{s} \times \boldsymbol{V}_{\oplus}))}{\textrm{c}^{2}} =
    -\frac{\boldsymbol{b}\cdot(\boldsymbol{s} + \Delta\boldsymbol{s})}{\textrm{c}}
\end{aligned}
,\end{equation}

where the second term represents the correction for the annual aberration

\begin{equation}\label{corr_main_aberr}
    \Delta\boldsymbol{s}=-\frac{(\boldsymbol{s}\times(\boldsymbol{s} \times \boldsymbol{V}_{\oplus}))}{\textrm{c}} .
\end{equation}

\citet{Titov11a} showed that the Galactocentric acceleration vector $\boldsymbol{a}$ can be added to the conventional equation~(\ref{groupdelay_gcrs}) by replacing the barycentric velocity $\boldsymbol{V}_{\oplus}$ with the sum $\boldsymbol{V}_{\oplus} + \boldsymbol{a} \Delta t$,  where $\Delta t$  is the time since the reference epoch 2000.0. As a result the corresponding correction for the aberration effect is

\begin{equation}\label{tau_aberr}
\begin{aligned}
\tau_{aberr}=&
\frac{\boldsymbol{b}\cdot(\boldsymbol{s}(\boldsymbol{s}\cdot(\boldsymbol{V}_{\oplus}+\boldsymbol{a} \Delta t)) -(\boldsymbol{V}_{\oplus}+\boldsymbol{a} \Delta t))}{\textrm{c}^{2}} = \\
=& \frac{\boldsymbol{b}\cdot(\boldsymbol{s}(\boldsymbol{s}\cdot\boldsymbol{V}_{\oplus})-\boldsymbol{V}_{\oplus})}{\textrm{c}^{2}} +
\Bigg(\frac{(\boldsymbol{b}\cdot\boldsymbol{s})}{\textrm{c}} \frac{(\boldsymbol{s}\cdot\boldsymbol{a}) \Delta t}{\textrm{c}} -
\frac{(\boldsymbol{b}\cdot\boldsymbol{a}) \Delta t}{\textrm{c}^2}\Bigg) =\\
=& \tau_{aberr_1} + \tau_{aberr_2}.
\end{aligned}
\end{equation}

Since the proper motions affect the source positions in the form $\boldsymbol{s'} = \boldsymbol{s} + \Delta\boldsymbol{s} = \boldsymbol{s} + \boldsymbol{\mu} \Delta t$, the vector of proper motion should be represented as a sum of two components $\boldsymbol{\mu} = \boldsymbol{\mu_1} + \boldsymbol{\mu_2} = \boldsymbol{s} (\boldsymbol{s} \cdot \boldsymbol{\mu}) + \boldsymbol{\mu_2}$, where $\boldsymbol{\mu_2}$ is the component along the acceleration vector $\boldsymbol{a}$, and $\boldsymbol{\mu_1}$  has the direction perpendicular to vectors $\boldsymbol{a}$ and  $\boldsymbol{s}$. Then, the additional delay induced by random-like apparent proper motion of a radio source is given by

\begin{equation}\label{tau_pm}
\tau' = -\frac{(\boldsymbol{b}\cdot\boldsymbol{\mu}) \Delta t}{\textrm{c}} = -\frac{(\boldsymbol{b}\cdot\boldsymbol{s})(\boldsymbol{s}\cdot\boldsymbol{\mu}) \Delta t}{\textrm{c}} - \frac{(\boldsymbol{b}\cdot\boldsymbol{\mu_2}) \Delta t}{\textrm{c}} .
\end{equation}

It should be noted that even if the projection of $\Delta\boldsymbol{s}$ from equation~(\ref{corr_main_aberr}) on the vector $\boldsymbol{s}$ is zero, i.e. $\boldsymbol{s} \cdot \Delta\boldsymbol{s} = 0$, the time delay in equations~(\ref{tau_aberr}) and~(\ref{tau_pm}) describes a projection of $\Delta\boldsymbol{s}$ on the baseline vector $\boldsymbol{b}$. Therefore, in a general case, the delays in equations~(\ref{tau_aberr}) and~(\ref{tau_pm}) as well as both components of the equation~(\ref{tau_aberr}), $\tau_{aberr_1}$ and $\tau_{aberr_2}$, are non-zero values.

From the sum of equations~(\ref{tau_aberr}) and~(\ref{tau_pm}),

\begin{equation}\label{tau_aberr_plus_pm}
\begin{aligned}
\tau_{aberr_2}+\tau'  =& \frac{(\boldsymbol{b}\cdot\boldsymbol{s}) (\boldsymbol{s}\cdot\boldsymbol{a}) \Delta t}{\textrm{c}^2} -
\frac{(\boldsymbol{b}\cdot\boldsymbol{a}) \Delta t}{\textrm{c}^2} \\
& -\frac{(\boldsymbol{b}\cdot\boldsymbol{\mu_1}) \Delta t}{\textrm{c}} -
\frac{(\boldsymbol{b}\cdot\boldsymbol{\mu_2}) \Delta t}{\textrm{c}} = \\
 =&\frac{(\boldsymbol{b}\cdot\boldsymbol{s}) (\boldsymbol{s}(\boldsymbol{a} - \textrm{c}\boldsymbol{\mu})) \Delta t}{\textrm{c}^2} -
\frac{(\boldsymbol{b}\cdot(\boldsymbol{a}+\textrm{c}\boldsymbol{\mu_2})) \Delta t}{\textrm{c}^2}
\end{aligned}
,\end{equation}

it can be concluded that the first term, proportional to the geometric delay $\tau_{geom}= -\frac{\boldsymbol{b}\cdot\boldsymbol{s}}{\textrm{c}}$, is sensitive to the component $\boldsymbol{\mu_1}$ alone. This implies that the estimate of the Galactocentric acceleration with the first term of equation~(\ref{tau_aberr_plus_pm}) is likely to be more accurate than from analysis of individual proper motion of reference radio sources.

If the model (equation~(\ref{groupdelay_gcrs})) described the observations perfectly, then the scale factor (F) determined by

\begin{equation}\label{Fdef}
F = \frac{\tau_{group}}{\tau_{geom}}
\end{equation}

is equal to unity for all observations, i.e. $F \equiv 1$ (e.g.~\citet{MacMillan17}). However, if the proper motions are not estimated, then the first term in equation~(\ref{tau_aberr_plus_pm}) as a part of the group delay (equation~(\ref{groupdelay_gcrs})) is considered as a contribution to the scale factor. After division in equation~(\ref{Fdef}), the scale factor is given by

\begin{equation}\label{F}
F = 1+\frac{\boldsymbol{a}\cdot\boldsymbol{s}}{\textrm{c}} \Delta t = 1 + \Delta F
\end{equation}

and manifests itself as a variable parameter depending on the Galactocentric acceleration, radio source position, and the time since a reference epoch. Consequently, the partial derivative of the group delay w.r.t. the scale factor as a time-dependent effect is given by

\begin{equation}\label{dtdF}
\frac{\partial \tau_{group}} {\partial F} = -\frac{\boldsymbol{b}\cdot\boldsymbol{s}}{\textrm{c}} ,
\end{equation}

or it can be expressed as a time-independent effect in the following way:

\begin{equation}\label{dtdFdt}
\frac{\partial \tau_{group}} {\partial \Big(\frac{F}{\Delta t}\Big)} = -\frac{\boldsymbol{b}\cdot\boldsymbol{s}}{\textrm{c}} \Delta t .
\end{equation}

\begin{figure}[ht]
          \includegraphics[trim=0 150 15 0,clip, width=\hsize]{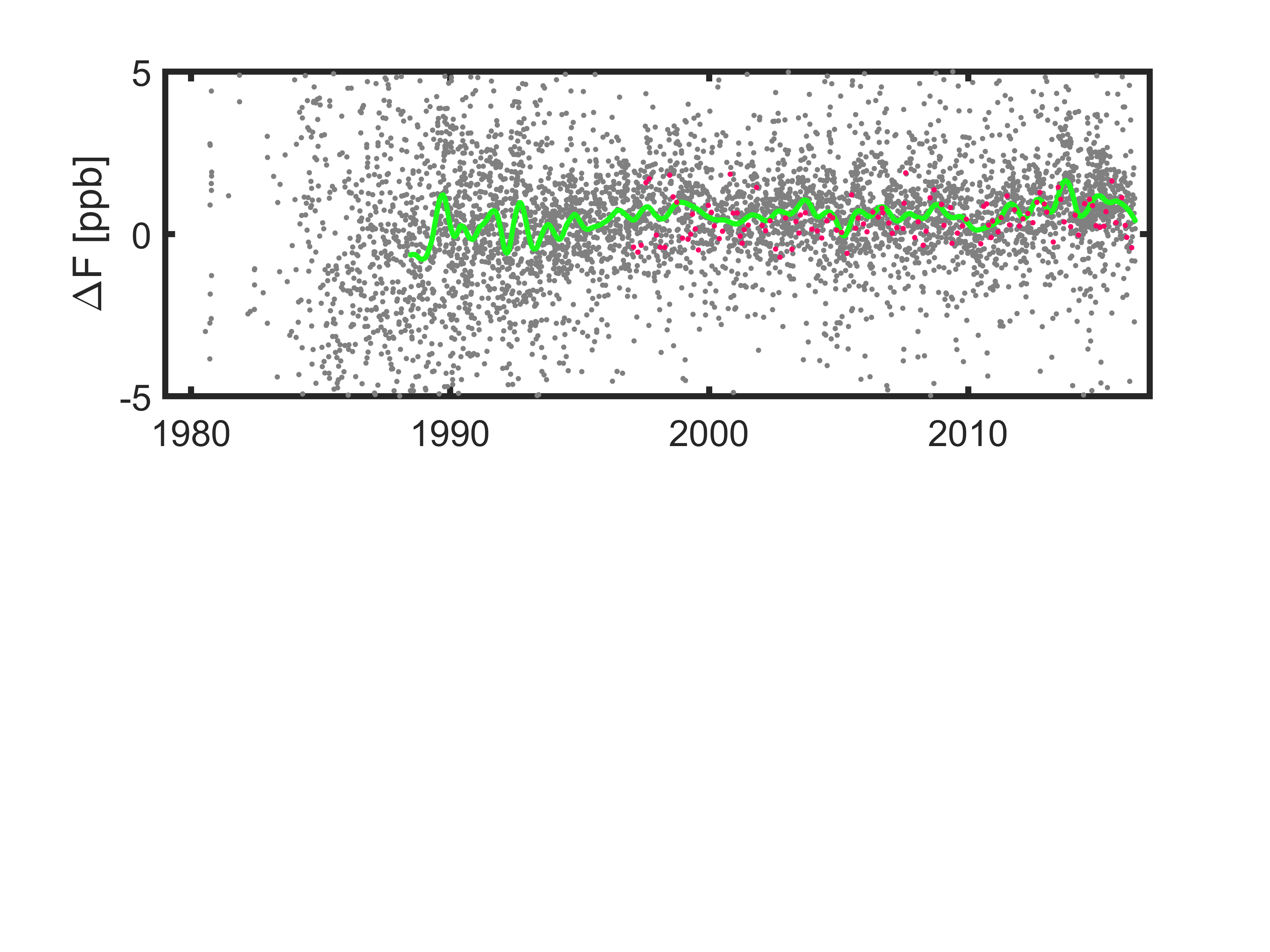}
  \caption{Session-wise corrections to the scale factor $\Delta F$ w.r.t. ITRF2014. The green line follows the smoothed estimates using the local least-squares regression. The magenta dots highlight the estimates from RDV sessions.}
     \label{Fig_scale_wrtITRF2014_all}
\end{figure}
\begin{figure}[ht]
     \includegraphics[trim=0 150 15 0,clip, width=\hsize]{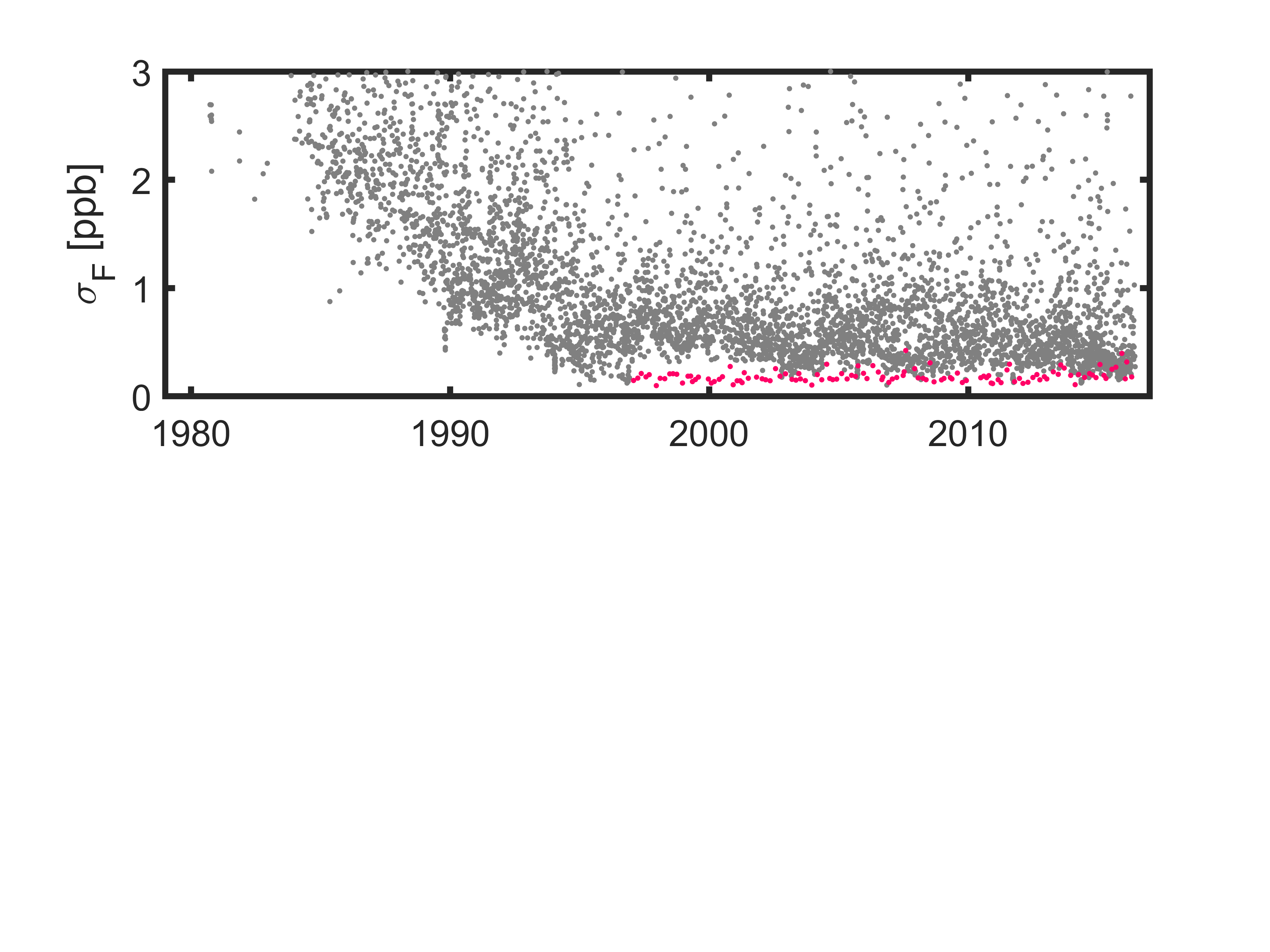}\\
\includegraphics[trim=0 150 15 0,clip, width=\hsize]{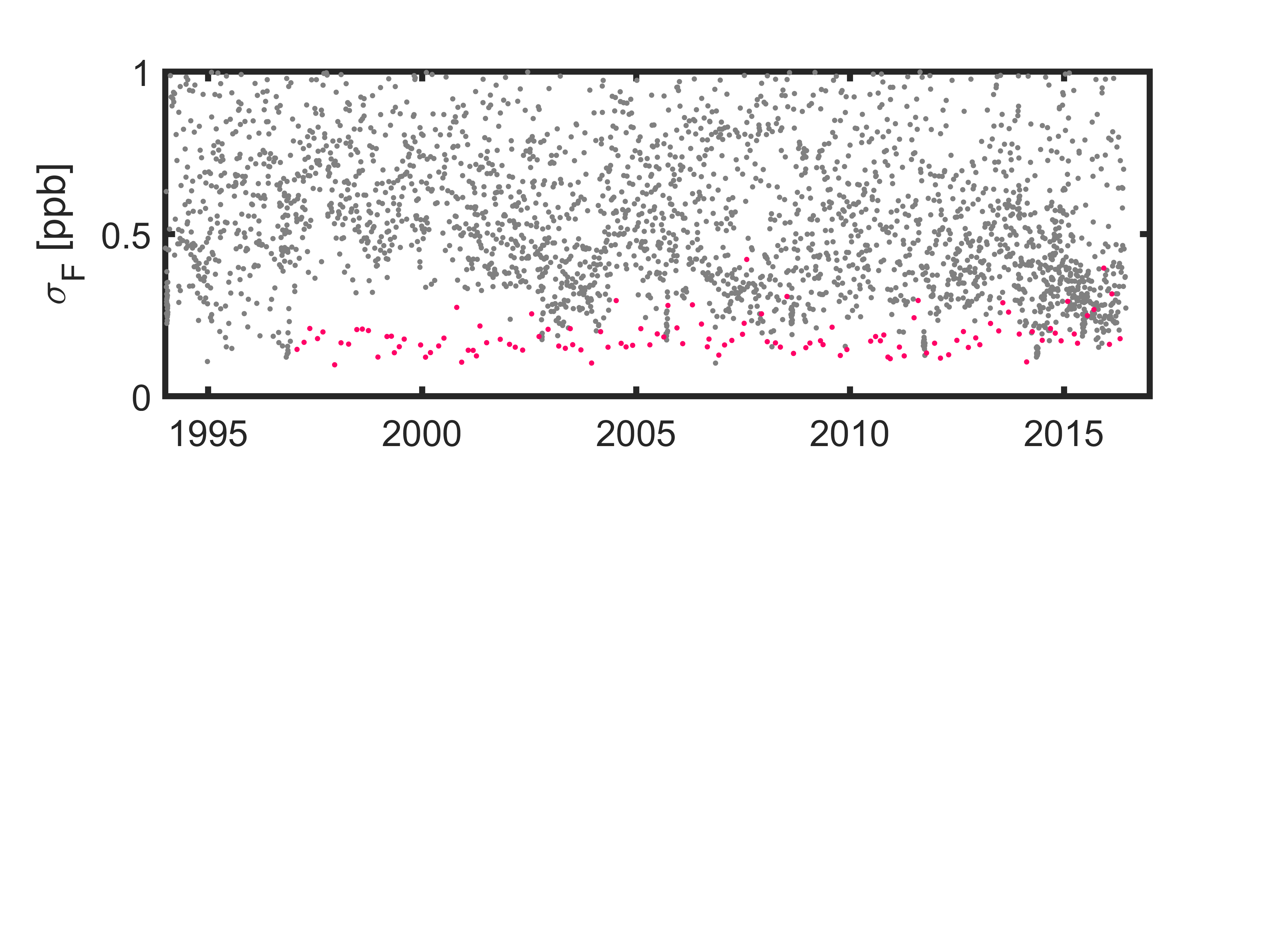}
  \caption{Formal errors of the session-wise corrections to the scale factor $\Delta F$ w.r.t. ITRF2014. The magenta dots highlight the formal errors from RDV sessions, shown in detail in the lower plot.}
     \label{Fig_sigma_scale_wrtITRF2014_all}
\end{figure}

\section{Analysis of VLBI data}
\label{Sec_analysis}
In this paper we analyse almost all VLBI sessions provided by the International VLBI Service for Geodesy and Astrometry (IVS; \citep{Schuh12}) together with the Very Long Baseline Array (VLBA) Calibrator Survey (VCS) observing sessions VCS1-6 \citep{Beasley02, Fomalont03, Petrov05, Petrov06, Petrov08, Kovalev07} and VCS-II~\citep{Gordon16}, i.e. 5825~observing sessions from 1979.7 until 2016.5. A dedicated observational project was started in Australia with Parkes~64~metre radio telescope and other Long Baseline Array (LBA) facilities since 2004 to enforce astrometry in the southern hemisphere.

The VLBI analysis was carried out with the software VieVS~\citep{Boehm12} and processing of our standard reference solution (P1) followed the International Earth Rotation and Reference Systems Service (IERS) Conventions~2010 \citep{iers10} and the technique specific VLBI analysis recommendations. Further details of the parametrisation and analysis set-up for our standard VLBI solution are given in \citet{Krasna15}.

\subsection{Scale factor}
The scale factor is usually not estimated in the routine procedure of geodetic VLBI data analysis. However, there is a possibility to estimate the scale factor either  as a single parameter (in a session-wise or global adjustment) or  as a specific parameter for each radio source individually within the global solution, similar to the radio source coordinates.

\subsubsection{Scale factor as a single parameter}
Using equation~(\ref{dtdF}) we obtained the session-wise estimates of the scale factor from the standard solution (Figure~\ref{Fig_scale_wrtITRF2014_all}), where the station positions and velocities were fixed to the a priori values in ITRF2014~\citep{Altamimi16}. The Earth orientation parameters and positions of the radio sources (no-net-rotation on the defining ICRF2~\citep{Fey09, Fey15} radio sources) were estimated in the usual way on a session-wise basis. The green line in the figure shows the smoothed estimates using the local least-squares regression. The annual periodicity visible in the scale factor estimates is predominantly due to the omitted hydrology loading corrections in the station coordinates, as shown by~\citet{MacMillan17}, and the weighted mean over the estimated scale corrections of 0.7~part-per-billion (ppb) is in accordance with the definition of the ITRF2014 scale, which is obtained by the arithmetic average of the scale of satellite laser ranging (SLR) and VLBI solutions. The scale discrepancy between these two combined solutions building the ITRF2014 is 1.37$\pm$0.10~ppb at epoch 2010.0~\citep{Altamimi16}. There is an ongoing discussion about the reason for the systematic difference in scale as determined from the two high accurate space geodetic techniques. Recently, \citet{Appleby16} pointed out that there seems to be a range bias in the case of SLR, which would be responsible for about the half of the obtained scale discrepancy.

Figure~\ref{Fig_sigma_scale_wrtITRF2014_all} shows the respective formal errors of the scale factor estimates presented in Figure~\ref{Fig_scale_wrtITRF2014_all}. A rapid decrease of the formal errors can be seen between the early VLBI years and 1995. The most accurate estimates of the scale factor belong to the RDV (Research and Development VLBA) sessions and these are highlighted with the magenta dots.

\begin{figure*}[t]
   \begin{minipage}[b]{0.5\linewidth}
     \centering
     \includegraphics[width=\hsize]{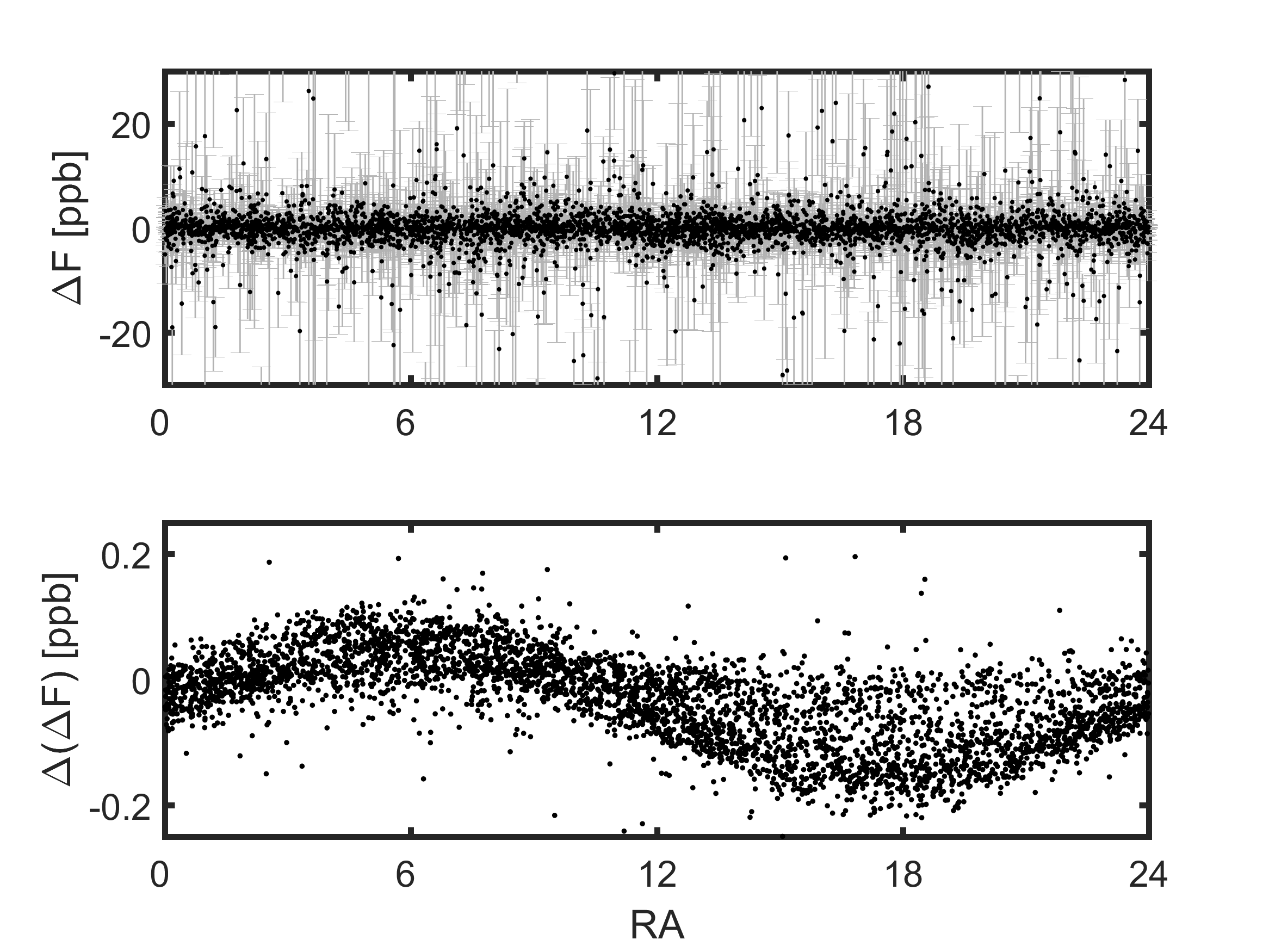}
   \end{minipage}
 \hspace{0.15cm}
   \begin{minipage}[b]{0.5\linewidth}
    \centering
    \includegraphics[width=\hsize]{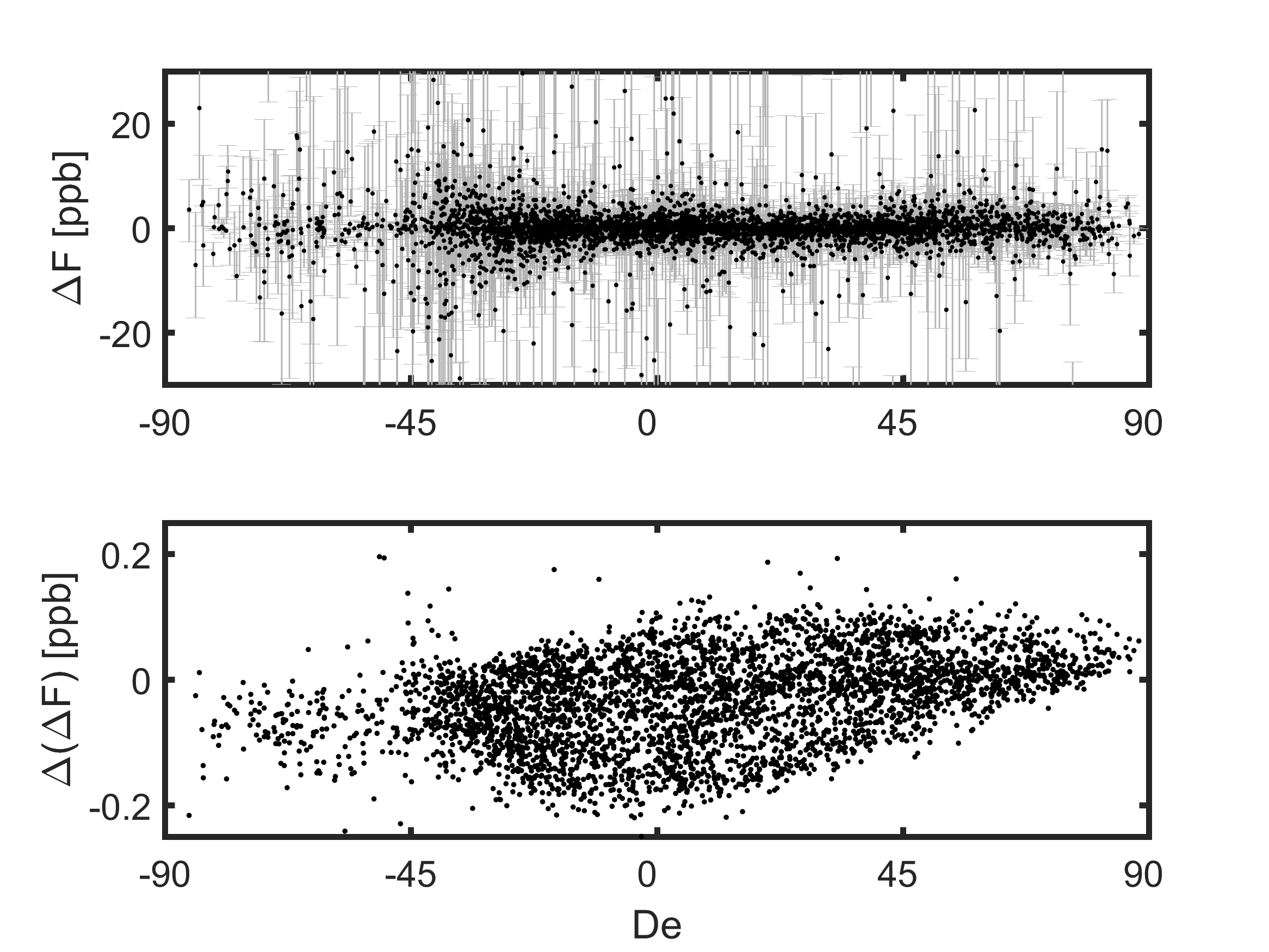}
  \end{minipage}
  \caption{Correction to the scale factor $\Delta F$ for all 4097 radio sources as a function of equatorial coordinates w.r.t. right ascension (left-hand side) and declination (right-hand side). Upper plots show the estimates from the parametrisation P1; lower plots depict the difference P1 minus P2.}
     \label{Fig_scale_S1S2_3obs}
\end{figure*}

\begin{figure*}[t]
   \begin{minipage}[b]{0.5\linewidth}
     \centering
     \includegraphics[width=\hsize]{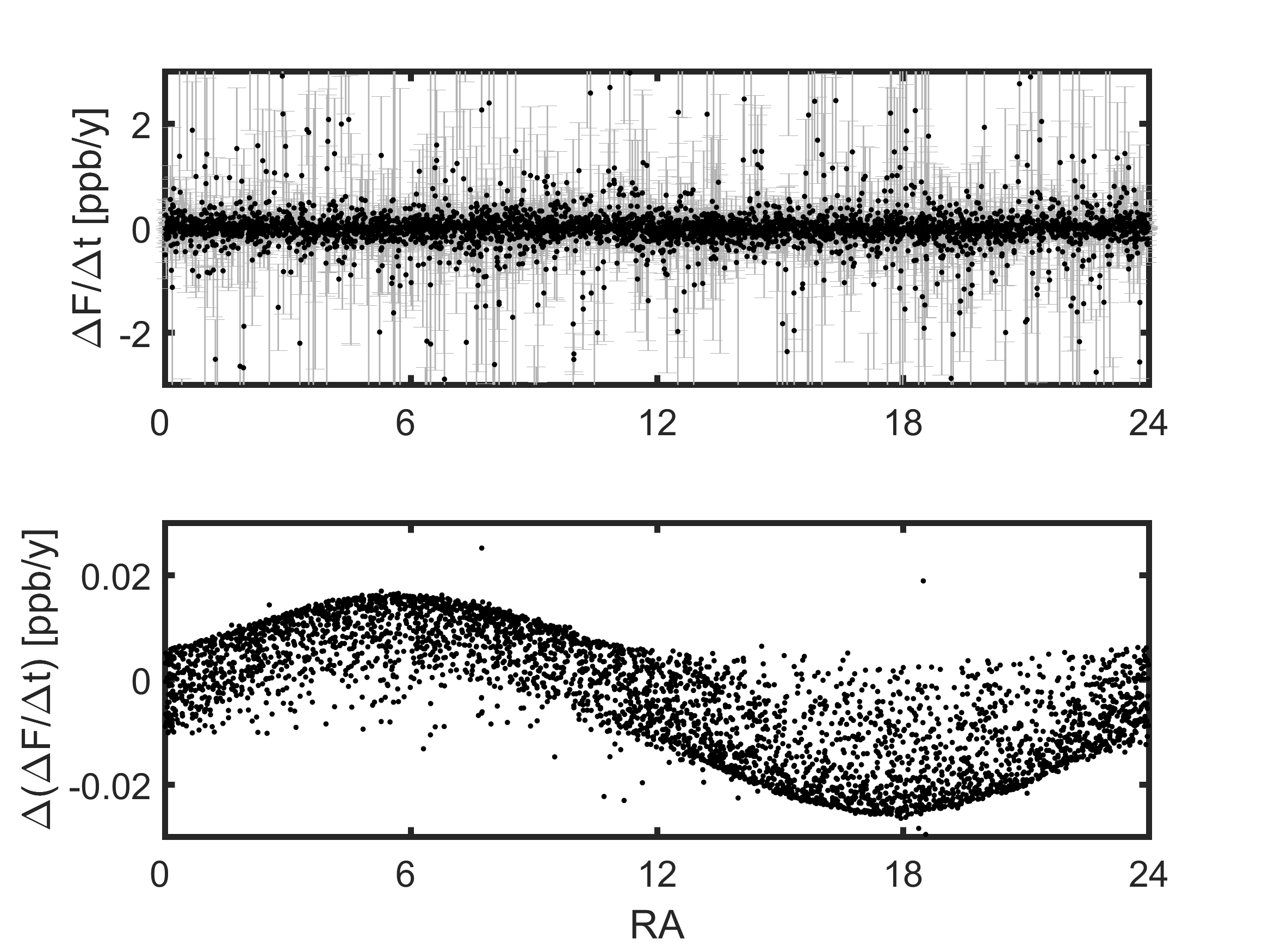}
   \end{minipage}
 \hspace{0.15cm}
   \begin{minipage}[b]{0.5\linewidth}
    \centering
    \includegraphics[width=\hsize]{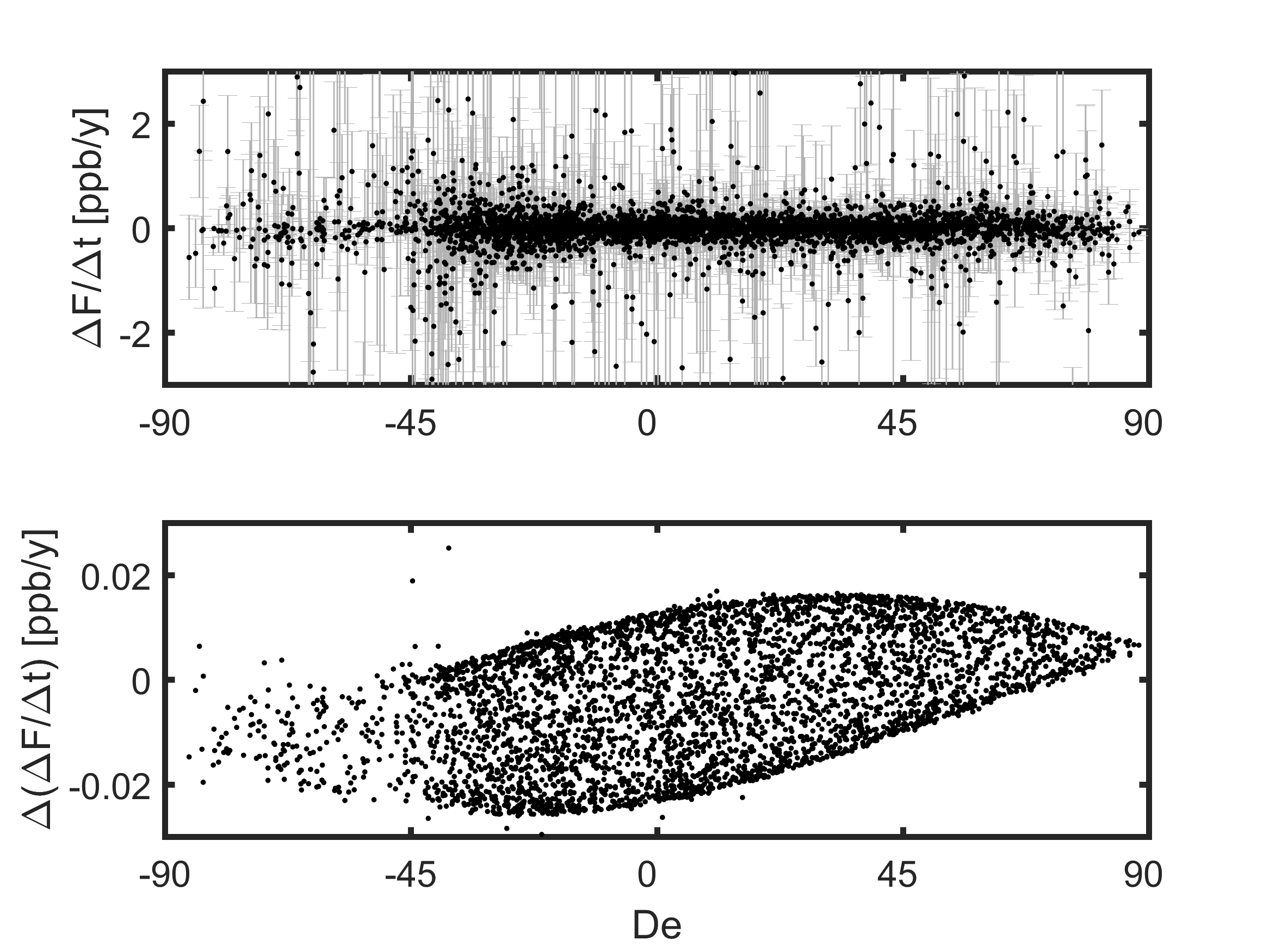}
  \end{minipage}
  \caption{Annual drift of the scale factor corrections $\Delta F/\Delta t$ for all 4097 radio sources as a function of equatorial coordinates w.r.t. right ascension (left-hand side) and declination (right-hand side). Upper plots show the estimates from the parametrisation P1; lower plots depict the difference P1 minus P2.}
     \label{Fig_scaleDt_S1S2_3obs}
\end{figure*}

\begin{table*}[t]
  \caption[]{Estimates and their formal errors of the dipole components in the $\Delta F/\Delta t$ a posteriori from the parametrisation P1. Different cut-off thresholds for number of radio source observations ($N$) applied.}
     \label{dipole}

\begin{tabular}{lrrrrrrrr}
        \hline
        \noalign{\smallskip}
$N$    &> 4 &   > 10 & > 50     & > 500 &       > 1 000 &> 10 000 & > 20 000  &> 50 000\\
        \noalign{\smallskip}
        \hline
        \noalign{\smallskip}
number of sources    &4062&     4001&3414       &573&476& 133 &87 &43\\
$A$ [\textmu as/yr] &   $7.1 \pm 0.2  $&$  8.2 \pm 0.3    $&$   5.2 \pm 0.2  $&$  5.1 \pm 0.3 $&$ 5.0 \pm 0.3 $&$4.8 \pm 0.4 $&$5.3 \pm 0.5 $&$ 4.6 \pm 0.7$\\
$\alpha_G$ [\degr]&     $281 \pm 3  $&$  281 \pm 3 $&$  281 \pm 3  $&$  281 \pm 4  $&$  280 \pm 5  $&$ 280 \pm 7  $&$  281 \pm 7  $&$  290 \pm 13$\\
$\delta_G$ [\degr]&   $-51 \pm 2 $&$ -55 \pm 2 $&$ -35  \pm 3 $&$ -34 \pm 3  $&$  -32 \pm 4 $&$ -28 \pm 5  $&$ -34  \pm 5  $&$ -24 \pm 8$\\
        \noalign{\smallskip}
        \hline
\end{tabular}
\end{table*}

\subsubsection{Scale factor as a specific global parameter for each radio source}
In this section we focus on the globally computed scale factor correction, that is from a common adjustment of the VLBI sessions, where we estimated this correction as a specific parameter for each radio source individually (except the 39~special handling sources). We ran a global solution with a standard parametrisation (P1), where corrections to the terrestrial reference frame (i.e. position and linear velocity of VLBI radio telescopes), celestial reference frame (i.e. position of radio sources), and the individual scale factor for each radio source are determined. Using equation~(\ref{dtdF}) for the partial derivative we obtained the scale factor corrections plotted in the upper part of Figure~\ref{Fig_scale_S1S2_3obs} as a function of right ascension (left-hand side) and as a function of declination (right-hand side). As a next step, the conventional equation for the group delay model \citep[chap.~11]{iers10} is extended by the Galactocentric acceleration vector $\boldsymbol{a}$ following \citet{Titov11a}. The SAD effect is reduced from the measurements a priori using the following values: $A~=~5$~\textmu as/yr, $\alpha_G = 267\degr$ and $\delta_G = -29\degr$. We refer to this parametrisation as to P2. Applying the P2 we ran a second global solution. The difference in $\Delta F$ (lower part of Figure~\ref{Fig_scale_S1S2_3obs}) between the standard solution and the solution with reduced SAD effect shows a clear systematic effect arising from the omitted SAD effect in the standard parametrisation. The maximum difference in the right ascension (left-hand side) and declination (right-hand side) coincides consequently with the direction to the Galactic centre and reaches about 0.2~ppb, whereas the weighted mean of the difference is $-0.016$~ppb.

To dispose of the time dependency of the scale factor we express the partial derivative of the group delay with equation~(\ref{dtdFdt}). Another two global solutions (with P1 and P2 parametrisation) were computed. The resulting annual drift of the scale factor corrections $\Delta F/\Delta t$ from P1 is plotted as a function of equatorial coordinates in the upper part of Figure~\ref{Fig_scaleDt_S1S2_3obs} and the lower part shows the difference between the P1 and P2 parametrisation. Comparing the Figures~\ref{Fig_scale_S1S2_3obs} and~\ref{Fig_scaleDt_S1S2_3obs} it is obvious that the scale factor corrections, expressed as a time-independent effect in the [ppb/yr] units, are free of large outliers. The annual variation of the scale factor due to the omitted SAD effect reaches about 0.02~ppb/yr, which reflects the mean observing time of about 10~years over all included radio sources.

Since we estimated the source-wise scale corrections together with the terrestrial (TRF) and celestial (CRF) reference frame within the global adjustment, a constant bias in the scale difference propagates also to the estimated terrestrial reference frame. Comparison of the TRF from the parametrisation P1 minus TRF from P2 in terms of the 14-parameters weighted Helmert transformation gives a negligible difference of $-0.011 \pm 0.001$~ppb ($-0.07 \pm 0.01$~mm) for the scale. It means that omitting the Galactocentric acceleration effect in the VLBI analysis does not have any significant effect on the global scale of the estimated TRF.

\subsection{Galactocentric acceleration vector}

\subsubsection{Galactocentric acceleration vector from source-wise scale factor}

The individual scale factor corrections $\Delta F/\Delta t$ could be fitted by the following model:

\begin{equation}\label{Ffitting}
F = a_1 \cos\alpha\cos\delta + a_2 \sin\alpha\cos\delta+ a_3 \sin\delta
\end{equation}

using the components of the acceleration vector $a_1, a_2, a_3$

\begin{equation}\label{a_component}
\begin{array}{lll}
a_1 = A \cos\alpha_G\cos\delta_G\\
a_2 = A \sin\alpha_G\cos\delta_G\\
a_3 = A \sin\delta_G
\end{array}
,\end{equation}

where $A$ is the amplitude of the Galactocentric acceleration and $\alpha_G$, $\delta_G$ are the equatorial coordinates of the vector direction. The three parameters determining the dipole ($a_1, a_2, a_3$) are to be estimated using the standard least-squares method. Then the amplitude $A$ and the coordinates $\alpha_G$, $\delta_G$ are given by

\begin{equation}\label{a_ampl}
A = \sqrt{a_1^2 + a_2^2 + a_3^2}
\end{equation}
\begin{equation}\label{a_direc}
\alpha_G = \arctan\Bigg(\frac{a_2}{a_1}\Bigg); \\
\delta_G = \arctan\Bigg(\frac{a_3}{\sqrt{a_1^2 + a_2^2}}\Bigg) .
\end{equation}

Table~\ref{dipole} shows the estimated dipole components from fitting the annual variation of the scale factor for different cut-off limits for the number of observations $N$ of each radio source in order to mitigate the effect of rarely observed radio sources and to verify the stability of the solution with respect to $N$. For a wide range of $N$ (from 50 to 20 000), where the compromise between the number of observations and the number of sources is fulfilled, the amplitude of the dipole effect varies between 4.8 and 5.3~\textmu as/yr, and the best formal error is 0.2 \textmu as/yr, i.e. much less than from the analysis of proper motions \citep{Titov11, Titov13, Titov16}. One can conclude that this approach is less sensitive to the peculiar behaviour of the active galactic nuclei. The estimated direction in declination (from $-28\degr$ to $-35\degr$) gets close to the value estimated by \citet{Reid04} within its formal error, whereas there is a offset of 14\degr between the estimated direction in right ascension ($\sim281\degr$) and the value 267\degr reported by \citet{Reid04}. The bias can be explained by the apparent proper motion of astrometrically unstable radio sources. This instability is caused by internal motion of matter in relativistic jet.

As shown in Figure~\ref{Fig_scale_wrtITRF2014_all} the time series of the scale factor estimates contains an annual signal originating predominantly from the omitted hydrology loading corrections. Therefore, we computed another solution based on P1, where we applied the hydrology loading displacement time series provided by the NASA Goddard Space Flight Center VLBI group~\citep{Eriksson14} on the station coordinates a priori. We ran again a global solution where the annual drift of the individual scale factor corrections was estimated in the same way as in the solutions mentioned above. From fitting the estimates belonging to radio sources with more than 50~observations, the following dipole components were obtained: $A = 5.1~\pm~0.2$~\textmu as/yr, $\alpha_G = 279\degr~\pm~3\degr$ and $\delta_G = -35\degr~\pm~3\degr$. This yields a difference of 0.1~\textmu as/yr in $A$ and 2\degr in $\alpha_G$ with respect to the solution based on parametrisation P1, which lies within the formal errors of the estimates and shows that the seasonal hydrology variation in the station coordinates does not have any significant effect on the estimated dipole components.

\subsubsection{Galactocentric acceleration vector as a global parameter}
The dipole components can be also estimated as global parameters in a VLBI solution in the form of the Galactocentric acceleration vector. The parametrisation P2 allowed us to build the partial derivative of the group delay with respect to the three components ($a_1, a_2, a_3$) of $\boldsymbol{a}$. We ran a global solution in which corrections to the terrestrial reference frame (i.e. position and linear velocity of VLBI radio telescopes), celestial reference frame (i.e. position of radio sources), and the three components of the Galactocentric acceleration vector were determined. Table~\ref{tab:GAglobres} summarises the estimated SAD parameters given with the amplitude $A$ of the Galactocentric acceleration vector and with the direction in $\alpha_G$ and $\delta_G$. The first column shows the resulting Galactocentric acceleration vector from a global adjustment of all VLBI sessions in our dataset (i.e. from 1979.7 until 2016.5). The second column contains the Galactocentric acceleration vector determined within an adjustment of large global networks only, represented by the following specific IVS programmes: the National Earth Orientation Service (NEOS-A) sessions, Rapid turnaround IVS-R1 and IVS-R4 sessions, and all available two-weeks CONTinuous campaigns (CONT), i.e. $\sim$2000 sessions from 1993 until 2016.5. Both solutions yield a value for the amplitude $A$, which is consistent with the estimates published in the last few years (\citet{Titov11, Xu12, Xu13, Titov13, Titov16, MacMillan14}). The direction in declination of the vector coincides with the value estimated by \citet{Reid04} if only the large network sessions were included in the solution, which implies that this procedure is sensitive to the inclusion of weak networks.

\begin{table}
  \caption[]{Galactocentric acceleration vector estimated as global parameter within VLBI solutions.}
     \label{tab:GAglobres}
     \setlength{\tabcolsep}{10pt}
\begin{tabular}{l r r}
        \hline
        \noalign{\smallskip}
  & 1979.7 - 2016.5 & 1993.0 - 2016.5 \\
  & $\sim$5800 sessions & $\sim$2000 sessions\\
        \noalign{\smallskip}
        \hline
        \noalign{\smallskip}
$A$ [\textmu as/yr] & $6.1 \pm 0.2 $ & $5.4 \pm 0.4 $\\
$\alpha_G$ [\degr]& $260 \pm 2 $  & $273 \pm 4 $\\
$\delta_G$ [\degr]& $-18 \pm 4 $  & $-27\pm 8 $\\
        \noalign{\smallskip}
        \hline
\end{tabular}
\end{table}

\section{Conclusions}
We introduce a new method to detect the secular aberration drift induced by the Galactocentric acceleration from the geodetic VLBI measurements. It is based on fitting the scale factor corrections estimated for each source individually within a global solution. In \citet{Titov13} we used the individual proper motion of reference radio sources to visualise the dipole effect caused by the Galactocentric acceleration. However, because of contamination by large apparent motions induced by relativistic jet, the proper motions have large random noise almost hiding the dipole systematics. However, if the dipole components in proper motion are not estimated, the systematic effect comes to the scale effect that, in contrast to the proper motion, is free of the relativistic jet problems.

From fitting the individual scale factor corrections of sources with more than 50~observations during  1979.7 - 2016.5 we obtained a Galactocentric acceleration vector with an amplitude of 5.2 $\pm$ 0.2~\textmu as/yr and direction $\alpha_G = 281\degr \pm 3\degr$ and $\delta_G = -35\degr \pm 3\degr$. The SAD was also estimated directly within a global adjustment of the VLBI data. The Galactocentric acceleration vector determined from the selected large network IVS sessions after 1993 ($A =  5.4 \pm 0.4$~\textmu as/yr, $\alpha_G = 273\degr \pm 4\degr$, $\delta_G = -27\degr \pm 8\degr$) is closer to the value reported by  recent papers (e.g. \citet{Reid04} and \citet{deGrijs16}) than the estimate from the entire VLBI history. The formal error of the amplitude of the acceleration vector estimated by means of the scale factor is about five times better than the estimate from the proper motions \citep{Titov16} and comparable to the accuracy expected to be obtained with the Gaia mission \citep{Mignard12}. This demonstrates the advantage of the new approach with respect to the traditional procedure.

Our result also points out the potential impact of neglecting the Galactocentric acceleration on the scale factor and, finally, on the geodetic positions of the radio telescopes. The scale factor becomes dependent on the positions of the reference radio sources, i.e. larger in the direction of the Galactic anti-centre, and smaller in the direction of the Galactic centre. In a global sense, the quality of the geodetic results is worsening and, moreover, the loss of quality would increase linearly with time.
The results presented in this paper were computed with the VieVS software and verified with the OCCAM software package.

\begin{acknowledgements}
The authors thank the anonymous referee for her/his constructive comments. The authors acknowledge the IVS and all its components for providing VLBI data \citep{Nothnagel15}. The Long Baseline Array is part of the Australia Telescope National Facility which is funded by the Australian Government for operation as a National Facility managed by CSIRO (P483 and V515). This study also made use of data collected through the AuScope initiative. AuScope Ltd is funded under the National Collaborative Research Infrastructure Strategy (NCRIS), an Australian Commonwealth Government Programme. The Very Long Baseline Array (VLBA) is operated by the National Radio Astronomy Observatory, which is a facility of the National Science Foundation, and operated under cooperative agreement by Associated Universities, Inc. Hana Kr{\'a}sn{\'a} works within the Hertha Firnberg position T697-N29, funded by the Austrian Science Fund (FWF). This paper has been published with the permission of the Geoscience Australia CEO.
\end{acknowledgements}

\bibliographystyle{aa} 
\bibliography{references_Titov_Krasna_AA2017} 

\end{document}